\documentclass[12pt]{amsart}
\usepackage[russian,english]{babel}
\usepackage{amsfonts,amssymb,amsthm}
\usepackage[mathscr]{eucal}
\usepackage{hyperref}
\usepackage{times}
\usepackage{color}

\title{Stable interactions via proper deformations}
\author{D. S. Kaparulin, S. L. Lyakhovich and A. A. Sharapov}
\address{Physics Faculty, Tomsk State University, Tomsk 634050, Russia}
\email{dsc@phys.tsu.ru, sll@phys.tsu.ru, sharapov@phys.tsu.ru}

\thanks{The work was partially supported by the Tomsk State
University Competitiveness Improvement Program. DSK and AAS were partially supported by
the RFBR grant 13-02-00551. SLL was supported in part by the RFBR
grant 14-01-00489.}

\begin{document}

\maketitle

\begin{abstract}

A new method is proposed for switching on interactions that are compatible
  with global symmetries and conservation laws of the original free theory.
  The method is applied to the control of stability in Lagrangian and non-Lagrangian theories with higher derivatives.
  By way of illustration,  a wide class of stable interactions is constructed for the Pais-Uhlenbeck oscillator.

\end{abstract}

\section{Introduction}

The inclusion of consistent interactions  is a notorious problem in
various areas of field theory. The problem has several aspects
related to the notion  of consistency. In gauge theories, for
instance, consistency is usually understood as the requirement that
the theory still has the same number of gauge symmetries as it has
had before inclusion of interaction. This requirement is necessary
(but not always sufficient) to ensure that the free and interacting
models possess the same number of physical degrees of freedom.
Nowadays, the BRST theory provides the most powerful approach to the
control of gauge symmetries upon switching on interaction
\cite{BH1993,H1998,HK1997,BBH2000}. A complete control over physical
degrees of freedom is achieved in the involutive form of dynamics.
Using the concept of involution, a covariant perturbative procedure
for inclusion of interaction was proposed in \cite{KLS-JHEP13}.
Apart from gauge symmetries, the procedure accounts for hidden
integrability conditions (constraints) making no distinction between
Lagrangian and non-Lagrangian theories.

The stability of nonlinear dynamics is another crucial property of
interaction. Being understood as the boundedness of solutions to the
classical equations of motion, it provides a sufficient condition
for the existence of a stable quantum theory with a well-defined
vacuum state. This relationship between the classical and quantum
stability is almost obvious in theories without higher derivatives.
Once the energy is bounded\footnote{By a bounded function we mean a
function satisfying the two requirements: (i) the
function is bounded from below \textit{and} (ii) the level surfaces of the function
are bounded in the ambient space.}, the theory is stable
because each classical trajectory lies on a bounded isoenergetic
surface in the phase space and the quantum vacuum can be defined as
the state with the lowest energy. The sufficient stability
condition, however, becomes an issue in the higher-derivative
theories, where the canonical energy is usually unbounded even in
the linear approximation. For an introductory discussion of the
stability problem in higher-derivative theories we refer the reader
to \cite{Woodard-15}. The simplest example of higher-derivative
dynamics is provided by the Pais-Uhlenbeck oscillator. The stability
of this model at the free and interacting levels has been the
subject of numerous studies, see e.g. \cite{Llosa-PRA03, Smilga2005,
Nesterenko, Chen2013,Pavsic2013} and references therein.

In the recent papers \cite{KLS2014EurPhysC,KL-RFJ2014}, a new
non-perturbative approach to the stability of higher-derivative
systems has been proposed. The key ingredient of the approach is the
concept of Lagrange structure \cite{KLS-JHEP05,LS-JHEP06,LS-JHEP07}.
The role of  the Lagrange structure is twofold. On the one hand, it
makes possible a consistent quantization of a classical system even
though the classical equations of motion are non-Lagrangian, on the
other hand each Lagrange structure defines a specific correspondence
between symmetries and conservation laws of the theory. The latter
property can be viewed as an extension of the Noether theorem to the
non-Lagrangian theories \cite{KLS-JMP10}. Once the classical
equations of motion admit a bounded integral of motion, and the
Lagrange structure that relates this integral to the
time-translation symmetry, the theory can retain stability at the
quantum level. The bounded integral of motion, being connected with
the time translation, is naturally identified with a physical energy
of the system, which may differ from the canonical energy. The
surprising thing is that such integrals of motion and Lagrange
structures exist almost for any free  theory. Furthermore, their
number increases (in some precise sense) with increasing the order
of equations of motion. Upon inclusion of interaction  the equations
of motion and the Lagrange structure should be deformed in such a
way as to keep the bounded conservation law connected to the time
translation. In the case that the operator of free higher-order
equations admits factorization into the product of coprime,
lower-order operators, particular solutions to this deformation
problem were proposed in \cite{KLS2014EurPhysC,KL-RFJ2014,KKL2015}.

In the present work, we construct a new class of stable nonlinear
theories with higher derivatives, where the interaction is
introduced by the proper deformation method proposed in
\cite{KLS-JMP10}. Contrary to the method of \cite{KLS2014EurPhysC},
the proper deformation of classical equations does not change the
Lagrange structure, deforming only the characteristic of the
conservation law. Under certain conditions the deformed equations describe a stable dynamical system,  at least in
some vicinity of classical vacuum. It is significant that the resulting non-linear theory is always Hamiltonian, even though the interaction vertices introduced by proper deformation are generally non-Lagrangian.

The paper is organized as follows. In Sec. 2, we recall the
definition of the Lagrange structure and the formulation of the
generalized Noether theorem for not necessarily Lagrangian theories.
The main results of the paper are contained in Sec. 3. Here, after
explaining the notion of a proper deformation, a set of conditions
is specified whereby a given symmetry of equations of motion remains
unchanged, while the corresponding integral of motion gets linear
and quadratic corrections in deformation. In Sec. 4, we apply the
proper deformation technique to the translation-invariant theories
and construct a class of stable non-linear systems with higher
derivatives. The general method is then illustrated by the example
of Pais-Uhlenbeck oscillator. In the concluding Sec. 5 we summarize
the results.

\section{Conservation Laws, Symmetries and Lagrange structures}

We start with a brief review of the concept of a Lagrange structure
and its relation to the symmetries and conservation laws. To
simplify formulas below  we  use the condensed index notation
\cite{deWitt}. According to this notation, the set of fields
$\{\varphi^i\}$ is labeled by the index $i$ that includes all
discrete indices as well as local coordinates $x^{\mu}$ on a
$d$-dimensional spacetime manifold $M$. As usual, summation over
repeated condensed indices implies integration over the spacetime and the
partial derivatives $\partial_i=\partial/\partial\varphi^i$ are
understood as variational ones. The equation $S\doteq 0$ means that a local functional $S[\varphi]$ is given by an integral over $M$ of an exact $d$-form, $S=\int_M dJ$; and hence, $\partial_i S=0$.

The dynamics of fields are governed by a set of partial differential
equations
\begin{equation}\label{T}
T_{a}(\varphi)=0\,.
\end{equation}
Since we do not assume the equations of motion to come from the
least action principle, the (discrete part of) indices $a$ and $i$
may run over different sets. For Lagrangian theories $a=i$ and
$T_i=\partial_i S,$ with $S$ being a classical action. The set of
all solutions to (\ref{T}) is called the \textit{mass shell} or
simply \textit{shell}.

By definition, a \textit{conservation law} $J$ is given by  an
on-shell closed differential form on $M$ of degree $d-1$. Under the
standard regularity conditions on the equations of motion (\ref{T}) this is equivalent to
\begin{equation}\label{Q}
Q^{a}T_{a}=\int_{M} dJ\doteq 0\,.
\end{equation}
The set of coefficients $Q=\{Q^a\}$ defining the left hand side is known
as the \textit{characteristic} of the conservation law $J$.
Like $T$'s, the components of the form $J$ and the characteristic $Q$ are supposed to be  smooth functions of fields and their space-time derivatives.
It is known that modulo
some trivialities there is a one-to-one correspondence between the
conservation laws and characteristics \cite{KLS-JMP10,OlverPJ}.

A variational vector field $X=X^i\partial_i$ is called a \textit{symmetry} of
the equations of motion (\ref{T}) if it preserves the mass shell, i.e.,
\begin{equation}\label{X}
X T_a=U{}_{a}^{b}T_b
\end{equation}
for some structure functions  $U_a^b$. Two symmetries $X_1$
and $X_2$ are considered as equivalent if they coincide on
shell, i.e.,
$$
X_1-X_2=T_aK^{a}\,,\qquad
(U_1){}^b_a-(U_2){}^b_a=K^{b}T_a
$$
for some set of variational vector fields $K^a=K^{ai}\partial_i$. It should be emphasized that
for non-Lagrangian theories the symmetries and conservation laws are
not related by the Noether theorem anymore.

A set of variational vector fields $V_a=V_a^i\partial_i$ is said to define a \textit{Lagrange structure} for the equations of motion  (\ref{T}) if the following compatibility condition is
satisfied:
\begin{equation}\label{V}
V_aT_b-V_b T_a=C_{ab}^dT_d
\end{equation}
for some structure functions $C_{ab}^d=-C_{ba}^d$. The distribution $V=\{V_a\}$ is called the \textit{Lagrange anchor}.

In this paper, we are mostly interested in the so-called
\emph{strongly integrable Lagrange structures}. These satisfy the following additional
conditions:
\begin{equation}\label{VV}
[V_a, V_b]=C_{ab}^dV_d\,,\qquad V_a
C_{bd}^f+C_{ab}^{g}C_{dg}^f+cycle( a,b,d)=0\,.
\end{equation}
The first relation just says that the anchor distribution $V$ is
integrable. For linearly independent  $V$'s  the second relation is
then a mere consequence of the Jacobi identity for the Lie bracket
of vector fields. From the geometrical viewpoint, Rels. (\ref{VV})
define a Lie algebroid with anchor $V=\{V_a\}$ and Rel. (\ref{V})
can be regarded as the closedness condition for the Lie algebroid
one-form $T=\{T_a\}$. For a quick introduction to the theory of Lie
algebroids we refer the reader to \cite{Weinstein}.

Notice that each Lagrangian theory admits the \emph{canonical Lagrange anchor}
$\{V_i=\partial_i\}$ associated with the tangent Lie algebroid. In
that case, defining relation (\ref{V}) is automatically
satisfied due to the commutativity of variational derivatives,
$$
\partial_i T_j-\partial_i
T_i=\partial_i\partial_j S-\partial_j\partial_i S=0\,,\qquad
C_{ij}^k=0\,.
$$
It should be noted that the existence of a Lagrange structure
compatible with a given set of equations appears to be much less
restrictive condition for the dynamics than the existence of an
action. Many examples of non-Lagrangian equations together with
their Lagrange structures can be found in
\cite{KLS-JHEP05,LS3,KLS-IJMPA11, KLS-SIGMA12}. Let us stress that
the choice of a compatible Lagrange structure is not unique, and
even Lagrangian equations of motion may have a variety of different
(and hence, non-canonical) Lagrange structures. A particular example
of such a situation will be considered in Sec. \ref{application}.

In \cite{KLS-JMP10}, it was shown that each Lagrange anchor (be it
integrable or not) establishes a relationship between the conservation
laws and symmetries. Explicitly,
\begin{equation}\label{VQ}
X=Q^aV_a\,,
\end{equation}
where $Q$ is the characteristic of a conservation law (\ref{Q}).
Using (\ref{V}) one can easily see that
$$
X T_a=(C_{ba}^dQ^b-V_aQ^d)T_d\,.
$$
The symmetries of the form (\ref{VQ}) are called the
\emph{characteristic symmetries}.

We see that the correspondence between symmetries and conservation
laws is not given from the outset, but depends upon  the choice of a
Lagrange anchor. The classical Noether's theorem exploits the
canonical Lagrange anchor for Lagrangian equations of motion. In the
general case a dynamical system may admit several Lagrange anchors
leading to different relations between symmetries and conservation
laws. This means, in particular, that one and the same symmetry may
come from different conservation laws. For the translation-invariant
equations of motion, this allows one to construct a multi-parameter
family of Hamiltonians associated with various Lagrange anchors. In
some cases, this family  may contain a positive-definite Hamiltonian
even in the presence of higher derivatives. Some examples of this
kind can be found in
\cite{KLS2014EurPhysC,KL-RFJ2014,BK-AP05,DS-JPHYSA2006}.

\section{Proper deformations and conservation laws}

The second key ingredient of our construction, called the \textit{proper deformation}, was introduced in
\cite{KLS-JMP10}. This is defined as follows.  Let we have given two sets of equations of motion,
\begin{equation}\label{2T}
T_a(\varphi)=0 \quad \mbox{and}\quad  T'{}_a(\varphi)=0\,,
\end{equation}
for one and the same collection of fields $\{\varphi^i\}$ and let
$V$ be a strongly integrable Lagrange anchor for $T$'s. We say that the second set of equations is obtained by a  proper deformation of the first one if there exists a local functional
$\mathcal{S}$, called the \textit{generator of the proper deformation}, such that
\begin{equation}\label{PD}
T'{}_a=T_a+V_a\mathcal{S}\,.
\end{equation}
By making use of Rels. (\ref{VV}) one can easily see that the Lagrange anchor
$V$ is also compatible with the deformed equations,
so that both theories (\ref{2T}) share the same Lagrange structure. One can also regard Eq. (\ref{PD}) as
an  equivalence relation on the space of all
equations of motion compatible with a given Lagrange anchor $V$.
In general, the corresponding equivalence classes may be rather wide.
For example, any two Lagrangian theories are related by a proper
deformation w.r.t. the canonical Lagrange anchor:
$$
T'{}_i=T_i+\partial_{i}\mathcal{S}\,,\qquad T_i=\partial_iS\,,\qquad
T'{}_i=\partial_iS',\qquad \mathcal{S}=S'-S\,.
$$

Suppose now that the first theory in (\ref{2T}) has a symmetry generated by the
variational field $X$ which leaves invariant the generator of
proper deformation and the Lagrange anchor in the sense that
\begin{equation}\label{XS}
X\mathcal{S}\doteq 0\,,\qquad
[X, V_a]=U_a^bV_b\,,
\end{equation}
with $U$'s being given by (\ref{X}). Then, using  definition (\ref{V}), one can find
$$
X T'{}_a=U_a^b T'{}_b\,.
$$
The last relation tells us that $X$ is a symmetry of the deformed equations, too. Furthermore, if $X$ is a characteristic symmetry (\ref{VQ}) of the first
theory, then, under certain conditions to
be specified below, it remains so in the deformed  theory.

As was mentioned in the previous section, the generators
of symmetry are defined only modulo the equations of motion. Given a
characteristic symmetry, the general element of its equivalence
class reads
\begin{equation}\label{VQ+KT}
X=Q^aV_a+T_aK^{a}\,,
\end{equation}
with $K^a=K^{ai}\partial_i$ being some set of variational vector
fields. Let us further assume that $K$'s satisfy the relation
\begin{equation}\label{KV}
K^{ai}V_{a}^j+K^{aj}V_{a}^i=0\,.
\end{equation}
(More geometrically, the last condition can be written as $(V_a
\mathcal{S})(K^a\mathcal{S})\doteq 0$ for all local functionals
$\mathcal{S}$.) If $J$ is a conservation law of the first theory
with characteristic $Q$, then the deformed theory (\ref{PD})
possesses the conservation law $J'=J+J_1+J_2$, where the
$(d-1)$-forms $J_1$ and $J_2$ are defined by the relations
\begin{equation}\label{J12}
\int_M dJ_1=Q^a (V_a \mathcal{S})-T_a(K^{a}
\mathcal{S})\,,\qquad \int_M dJ_2=(K^a\mathcal{S})(V_{a}\mathcal{S}),.
\end{equation}
The characteristic $Q'$ of $J'$ is given by
\begin{equation}\label{Q1}
Q'{}^a=Q^a+K^{a}\mathcal{S}\,,\qquad \int
dJ'=Q'{}^aT'_{a}\,.
\end{equation}
It is easy to see that the symmetry associated with this
characteristic is equivalent to $X$,
$$
Q'^{a}V_a=X-T'_aK^{a}\,.
$$

We thus conclude that, under condition (\ref{KV}), each integrable Lagrange structure allows one to deform equations of motion together
with their conservation laws; in so doing, the deformed and
undeformed conservation laws correspond to essentially the same symmetry
transformation on the configuration space of fields. This
observation will be used to the control of stability in the next section.

\section{Application to the stability of interactions}\label{application}

In this section, we consider mechanical systems whose dynamics are
governed by ordinary differential equations (not necessarily
Lagrangian). We say that a system is \textit{classically stable} if
each its trajectory is bounded in the phase space. In particular,
this ensures the boundedness of motion in the configuration space.
It may happen that a classical system becomes stable when restricted
to some invariant domain in the phase space. Such a domain is
usually referred to as a stability island. Generally, it is not easy
to decide wether a given set of equations defines a stable system or
system with stability islands. In most cases the classical stability
is provided by an integral of motion whose level surfaces are
bounded in the phase space. If in addition the values of the
integral are bounded from below we call it bounded. For Lagrangian
theories without higher derivatives the role of such an integral is
often played by the canonical energy. Upon canonical quantization
the energy becomes a Hermitian operator with spectrum bounded from
below. This allows one to define the ground state as the state with
the smallest possible energy. It is the  existence of a ground (or
vacuum) state which is usually understood by the quantum stability.

Unfortunately, the energy argument above can not be applied directly
to the higher-derivative systems as the canonical Ostrogradsky's
energy of such systems is known to be unbounded, at least for regular
Lagrangians. This does not necessarily mean that the system has no
\textit{other} integrals of motion, some of which may happen to be
bounded as opposed to the canonical energy. Actually any bounded integral
ensures the classical stability and one can try to interpret it as
physical energy. To justify such an interpretation one only needs to
find a Lagrange structure which would relate this integral to the
time translation. On quantizing the theory by means of the
Lagrange structure, this bounded integral of motion should be identified with
the quantum Hamiltonian. By the correspondence principle one might
expect the spectrum of this Hamiltonian  to be bounded from below.

Finding of a bounded integral of motion for a given
higher-derivative system is quite a difficult  problem  in general.
The exception is provided by the linear higher-derivative systems,
where one can usually find a plenty of integrals of motion with the
desired property as well as Lagrangian structures linking them to
the time-translation symmetry. The proper deformation gives  a
simple method for constructing nonlinear theories with conserved
quantities related to the time translation. Whenever a linear model
admits a bounded integral of motion and Eqs. (\ref{XS}),
(\ref{VQ+KT}), (\ref{KV}) are satisfied, the conservation law $J'$
of the corresponding nonlinear theory is given by (\ref{Q}) and
(\ref{J12}) with
$X=-\dot{\varphi}^i\frac{\partial}{\partial\varphi^i}$. With a
suitable choice of $S$, the function $J'$ can be made bounded, so
that the nonlinear theory remains stable.

Let us now illustrate this general approach by the example of the
Pais-Uhlenbeck oscillator. The theory is described by the
forth-order differential equation
\begin{equation}\label{PU}
T(x)=\gamma(\stackrel{_{(4)}}{x}+(\omega_1^2+\omega_2^2)\ddot{x}+\omega^2_1\omega_2^2x)=0\,,\qquad
\stackrel{_{(k)}}{x}=\frac{d^kx}{dt^k}\,,\qquad
\gamma=\frac{1}{\omega^2_2-\omega_1^2}\,,
\end{equation}
where $0<\omega_1<\omega_2$ are the frequencies and $x(t)$ is a
single dynamical variable. We exclude the case of equal frequencies
as the corresponding motion is known to be unbounded due to the
phenomenon of resonance.

In \cite{KLS2014EurPhysC}, the following two-parameter families of
Lagrange anchors and characteristics were found:
\begin{equation}\label{V+Q}
V=-\gamma\Big(\Big(\frac{1}{\alpha}+\frac{1}{\beta}\Big)\frac{d^2}{dt^2}+
\Big(\frac{\omega_2^2}{\alpha}+\frac{\omega_1^2}{\beta}\Big)\Big)\,,\qquad
Q=\gamma((\alpha+\beta)\dddot{x}+(\alpha\omega_2^2+\omega_1^2)\dot{x})\,.
\end{equation}
For any nonzero constants $\alpha$ and $\beta$ they result to the
time-translation symmetry
\begin{equation}\label{X+K}
X\equiv-\dot{x}=V(Q)+K(T)\,, \qquad
K=-\gamma\frac{(\alpha+\beta)^2}{\alpha\beta}\frac{d}{dt}\,,
\end{equation}
Hereafter, to be more explicit we unfold our condensed notation
treating $V_a^i$ and $K^{ai}$ as integral kernels of differential
operators acting on the test function $\zeta(t)$,
$$
(V\zeta)(t)=\int dt' V(t,t')\zeta(t')\,,\qquad (K\zeta)(t)=\int
dt'K(t,t')\zeta(t')\,,
$$
where $t=i$ and $t'=a$ are pure continuous indices. The integral of
motion $J$ corresponding to the characteristic (\ref{V+Q}) reads
\begin{equation}\label{J0}
J=\frac{\gamma^2}{2} \big(\alpha(\dddot{x}+\omega_2^2\dot{x})^2+
\beta(\dddot{x}+\omega_1^2\dot{x})^2+\alpha\omega_1^2
(\dddot{x}+\omega_2^2\dot{x})^2+\beta\omega_2^2(\dddot{x}+\omega_1^2\dot{x})^2\big)\,.
\end{equation}
As is seen, the quadratic function $J$ is positive definite provided
that $\alpha,\beta>0$. Furthermore, the underlying quadratic form on
the phase-space of variables $x,\dot{x},\ddot{x},\dddot{x}$ is
nondegernerate whenever $\omega_1\neq\omega_2$, so that each level
surface of $J$ appears to be compact and the trajectories are
bounded. Notice that the canonical Ostrogradsky's energy corresponds
to $\alpha=-\beta$, in which case $V$ reduces to the canonical
Lagrange anchor for the Lagrangian  equations of motion. Being
intimately related with the time-translation, the conserved
quantity $J$ can be regarded as a physical energy of the
Pais-Uhlenbeck oscillator. This can be made positive by a proper
choice of the free parameters $\alpha$ and $\beta$.

Applying to (\ref{PU}) a proper deformation generated by a local
functional\footnote{The consistency of interaction \cite{KLS-JHEP13}
implies that both the linear and nonlinear theories have the same
number of degrees of freedom. This forces  us to restrict to the
ansatz (\ref{Sint}) without higher derivatives.}
\begin{equation}\label{Sint}
    \mathcal{S}=-\frac{1}{\gamma}\int dt U(x,\dot{x})\,,
\end{equation}
we get
\begin{equation}\label{Tint}
    \gamma(\stackrel{_{(4)}}{x}+(\omega_1^2+\omega_2^2)\ddot{x}+\omega^2_1\omega_2^2x)+
    \Big(\frac{1}{\alpha}+\frac{1}{\beta}\Big)\ddot{U}_x+
\Big(\frac{\omega_2^2}{\alpha}+\frac{\omega_1^2}{\beta}\Big)U_x=0\,,
\end{equation}
where $U_x$ denotes the Euler-Lagrange derivative of $\mathcal{S}$.
It is clear that the functional $\mathcal{S}$ is invariant under the
time translations.

The linear and quadratic corrections to the integral of motion
(\ref{J0}) due to the deformation are given by
\begin{equation}\notag\begin{array}{l}\displaystyle
J_1=\frac{1}{\gamma}U-\frac{1}{\gamma}\dot{x}\frac{\partial
U}{\partial \dot{x}}+
\gamma\frac{(\alpha+\beta)^2}{\alpha\beta}(\dddot{x}\dot{U}_x-\stackrel{_{(4)}}{x}U_x)+
\gamma\Big(\frac{\omega_1^2}{\alpha}+\frac{\omega_2^2}{\beta}\Big)(\alpha+\beta)(\dot{x}\dot{U}_x-\ddot{x}U_x)\,,\\[4mm]\displaystyle
J_2=-\frac{1}{2}\frac{(\alpha+\beta)^3}{\alpha^2\beta^2}
(2\ddot{U}_xU_x-\dot{U}_x^2)-\frac{1}{2}\frac{(\alpha+\beta)^2}{\alpha\beta}
\Big(\frac{\omega_2^2}{\alpha}+\frac{\omega_1^2}{\beta}\Big)U_x^2\,.
\end{array}\end{equation}
Excluding the fourth derivative in this expressions by means of
equation (\ref{Tint}), we can write the deformed integral of motion
in the form
\begin{equation}\label{Jint}\begin{array}{l}\displaystyle
J'=\frac{\gamma^2}{2}\Big\{
\alpha\Big(\dddot{x}+\omega_2^2\dot{x}+\frac{\alpha+\beta}{\alpha\beta\gamma}\dot{U}_x\Big)^2
+\beta\Big(\dddot{x}+\omega_1^2\dot{x}+\frac{\alpha+\beta}{\alpha\beta\gamma}\dot{U}_x\Big)^2+
\\[4mm]\displaystyle
+\alpha\omega_1^2\Big(\ddot{x}+\omega_2^2x+\frac{\alpha+\beta}{\alpha\beta\gamma}U_x\Big)^2+
\beta\omega_2^2\Big(\ddot{x}+\omega_1^2x+\frac{\alpha+\beta}{\alpha\beta\gamma}U_x\Big)^2\Big\}+\frac{1}{\gamma}U-\frac{1}{\gamma}\dot{x}\frac{\partial
U}{\partial \dot{x}}\,.
\end{array}\end{equation}
In general, this expression is not positive definite due to the
negative terms in the third line. However, if the interaction $U$ is
small enough the integral $J$ may well be bounded from below at
least in some neighborhood of zero in the phase-space. For example,
if we set $\alpha=\beta>0$, then $J'$ reduces to
\begin{equation}\label{1Jint}\begin{array}{l}\displaystyle
J'=\frac{\alpha\gamma^2}{2}\Big\{
\Big(\dddot{x}+\omega_2^2\dot{x}+\frac{2}{\alpha\gamma}\dot{U}_x\Big)^2
+\Big(\dddot{x}+\omega_1^2\dot{x}+\frac{2}{\alpha\gamma}\dot{U}_x\Big)^2+
\\[4mm]\displaystyle
+\omega_1^2\Big(\ddot{x}+\omega_2^2x+\frac{2}{\alpha\gamma}U_x\Big)^2+
\omega_2^2\Big(\ddot{x}+\omega_1^2x+\frac{2}{\alpha\gamma}U_x\Big)^2\Big\}+
\frac{1}{\gamma}U-\frac{1}{\gamma}\dot{x}\frac{\partial U}{\partial
\dot{x}}\,.
\end{array}\end{equation}
This expression is clearly positive on the whole phase space
whenever
$$
U-\dot{x}\frac{\partial U}{\partial \dot{x}}>0\,.
$$
The simplest possibility to satisfy this inequality is to take
$U=U(x)\geq 0$.

Notice that the equation describing the nonlinear Pais-Uhlenbeck
oscillator (\ref{Tint}) is non-Lagrangian, so that  the classical
Noether's theorem is not applicable to it anymore and the existence
of the conserved quantity $J'$ is not  a mere consequence of the
translation-invariance of the theory. However, applying the
techniques developed in \cite{S-IJMPA14, S-IJMPA15}, it is still
possible to bring  the nonlinear equation (\ref{Tint}) into the
Hamiltonian form with the function $J'$ playing the role of
Hamiltonian on the phase space of $x$, $\dot x$, $\ddot x$, $\dddot
x$. We are going to present the corresponding Poisson bracket
elsewhere.

\end{document}